# Obtaining material identification with cosmic ray radiography

Christopher Morris, Konstantin Borozdin, Jeffrey Bacon, Elliott Chen, Zarija Lukić, Edward Milner, Haruo Miyadera, and John Perry

*Los Alamos National Laboratory, Los Alamos, NM 87545, USA*

Dave Schwellenbach, Derek Aberle, Wendi Dreesen, J. Andrew Green, and George G. McDuff

*National Security Technologies, Los Alamos, NM 87544, USA*

Kanetada Nagamine

*KEK, Tsukuba, Ibaraki, 305-0801, Japan RIKEN, Wako, Saitama, 351-0198 Japan and UC-Riverside, CA 92521, USA*

Michael Sossong

*Decision Sciences, 14900 Conference Center Drive, Suite 125, Chantilly, VA 20151, USA*

Candace Spore and Nathan Toleman

*Department of Chemical and Nuclear Engineering, University of New Mexico, Albuquerque, NM 87131-0001, USA*

**Abstract.** The passage of muons through matter is mostly affected by their Coulomb interactions with electrons and nuclei. The muon interactions with electrons lead to continuous energy loss and stopping of muons, while their scattering off nuclei lead to angular "diffusion". By measuring both the number of stopped muons and angular changes in muon trajectories we can estimate density and identify materials. Here we demonstrate the material identification using data taken at Los Alamos with the Mini Muon Tracker.

## Introduction

Measuring the stopping of cosmic ray muons has been used for decades to radiograph objects such as pyramids and geological structures.[1-5] Technique based on measuring the multiple scattering of muons has been developed at Los Alamos more recently. This technique has been shown to be useful for locating materials with high atomic number in a contrast with a background of material of low atomic number.[6] The combination of energy loss and scattering has been suggested as a method to determine both material type and density (therefore providing material identification or MID) using focused beams of accelerator produced muons.[7] The combination of nuclear attenuation and Coulomb scattering has also been shown to provide MID in proton radiography.[8] In this paper we demonstrate MID, using the combination of stopping and scattering of cosmic ray muons. The data were taken at Los Alamos Neutron Science Center (LANSCE) with the Mini Muon Tracker shown in Figure 1. MMT was developed by a collaboration of Los Alamos National Laboratory, National Security Technologies, and Decision Sciences International Corporation. An example of multiple scattering radiographic image obtained with the MMT using overhead muons arriving from directions close to the zenith, is shown in Figure 2.

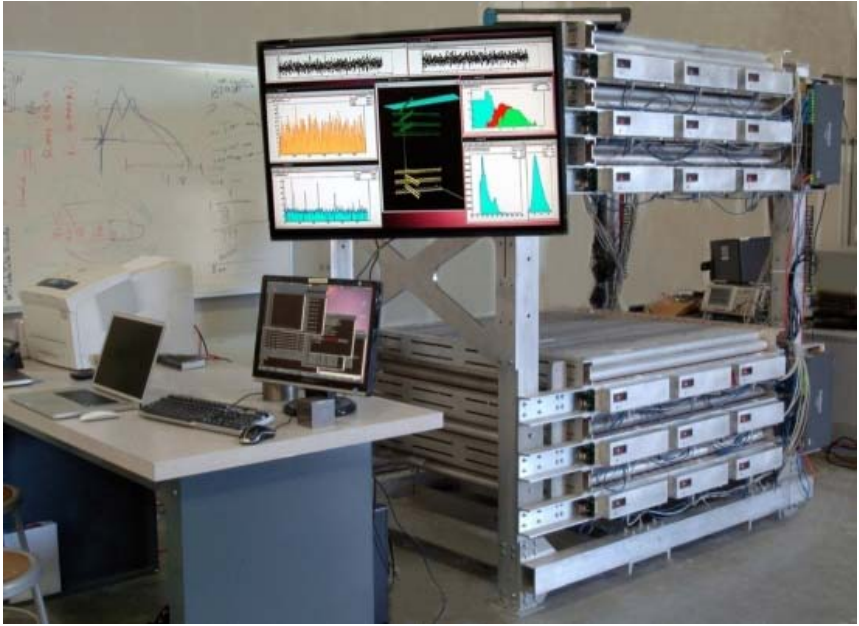

**Figure 1) Photograph of the MMT. Objects for study were placed in the approximately two-feet (60 cm) gap between the two detector "supermodules".**

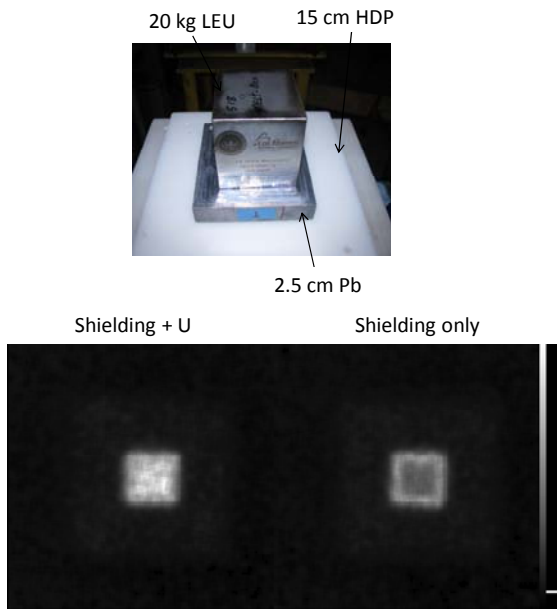

Figure 2) Top) A photograph of the bottom parts of a layered shielding box with a 10×10×10 cm$^3$ (~20 kg) uranium cube. When fully assembled, the shielding box surrounds the uranium with 5 cm of lead and surrounds the lead with 15 cm of borated polyethylene. Bottom right) a reconstructed image showing a slice at roughly the center of the object from a cosmic ray tomography of the shielding box. Bottom left) an image of the shielding box with the uranium. Grey scale of the image represents the strength of the scattering signal.

Both stopping and scattering measurements are being studied as potential tools to locate the fuel in the melted down reactor cores of the Fukushima reactors. The motivation for this work is the potential cost saving and reduction in human radiation dose for the three-decade long cleanup program at Fukushima Daiichi. MID may be also useful for homeland security and nuclear treaty applications.

In this paper we show how we identify materials combining the information from scattering and transmission of cosmic-ray muons.

## Cosmic ray imaging

Transmission (or stopping) imaging with cosmic rays is somewhat different from point source x-ray imaging in that both the intensity and the direction of the cosmic rays can be measured. The trajectory information can be used to generate a focused transmission image at any distance from the detector. Conceptually, the stopping length, $\lambda$, of cosmic rays in material is inversely proportional to the stopping rate and can be related to the energy spectrum, $dN(E)/dE$, as

$$\frac{1}{\lambda} = \frac{dN}{Ndx} = \frac{1}{N}\frac{dN}{dE}\frac{dE}{dx}.$$

A plot of the energy spectrum for overhead muons at sea level is shown in Figure 3. The energy loss, $dE/dx$, can be calculated using the Bethe-Bloch formula. Over a wide range of momentum, the energy loss for cosmic ray muons varies only logarithmically with momentum and is approximately proportional to the electron density, Z/A. For dense material, where $\lambda$ is short compared to the muon decay length, $l = \beta c \gamma \tau$, where $\beta$ and $\gamma$ are the usual kinematic quantities, c is the velocity of light, and $\tau = 2.2\ \mu s$ is the muon lifetime. The muon stopping can be understood as the shifting of the spectrum shown in

Figure 3 to the left, with the loss of muons with energies below some threshold.

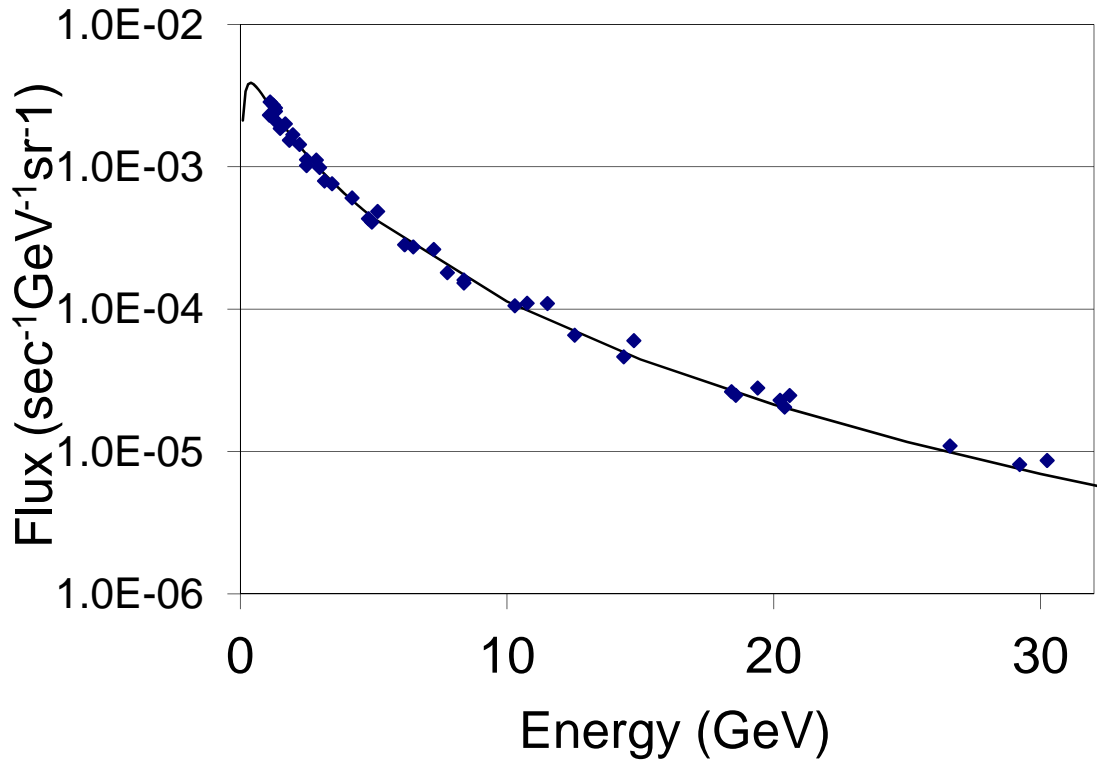

Figure 3) Spectrum of vertical cosmic ray flux at sea level. Solid symbols are the data.[9] The line is a parameterization.

We have measured the transmission through three thicknesses (each) of lead, concrete and steel. The transmission image was obtained by making a ratio of the image of transmitted cosmic rays with the target in place to an image with no target. The bottom and top tracks for transmitted trajectories had to intersect in a horizontal plane at the center of the object to within a radius of 1 cm. The resulting radiographs are shown in Figure 4.

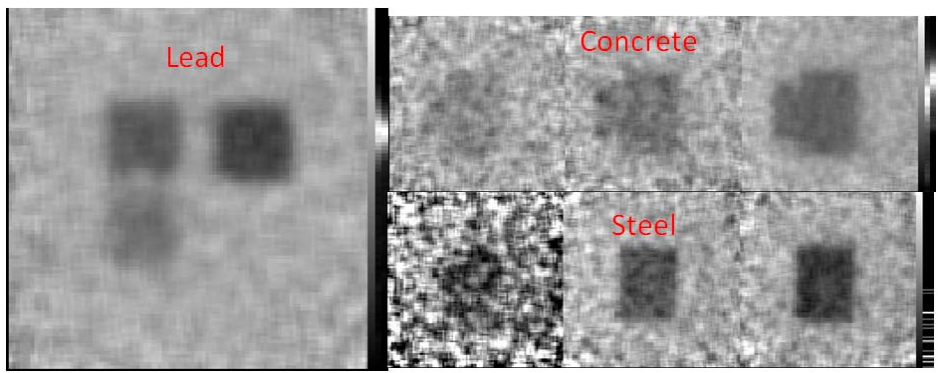

Figure 4) Transmission images for the lead (left), concrete (top right) and steel (bottom right) targets. The three thickness of lead were radiographed in a single run. The other targets were imaged during individual runs. The grey scale is linear and ranges in value from 0.6 (black) to 1.2 (white) in transmission for all targets.

The negative of the natural log transmission as a function of calculated energy loss is plotted in Figure 5. The slope of a linear curve constrained to go through the origin fitted to the data in Figure 5 is 1.03 GeV$^{-1}$. This is very close to the value of the peak of 1.2 GeV$^{-1}$ of the normalized spectrum, $\frac{1}{N}\frac{dN}{dE}$, in Los Alamos. The scatter between the different materials is not fully understood, but may be caused by the different geometry of the objects (concrete blocks being much thicker than lead). There is also a considerable uncertainty in the composition of the concrete, while the steel and the lead should be well understood.

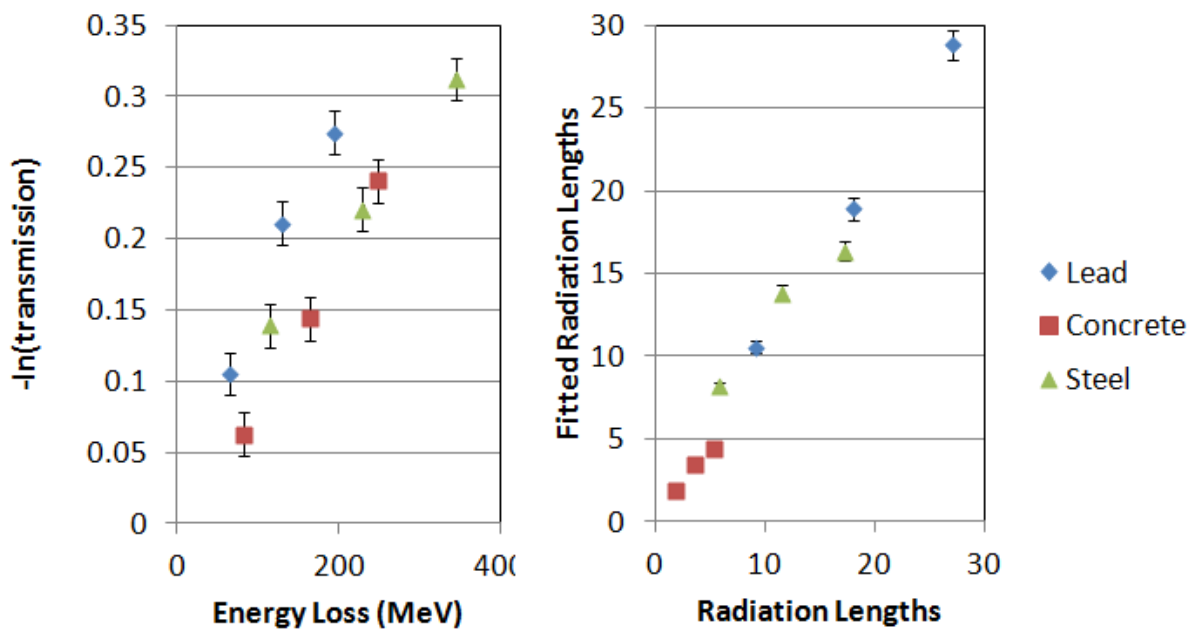

Figure 5) Left) Negative of the natural log of transmission vs. calculated energy loss for three thickness each of lead, concrete and steel. Right) Fitted radiation lengths vs. radiation lengths.

## Multiple scattering radiography

In addition to energy loss and stopping, cosmic rays also undergo Coulomb scattering from the charged atomic nuclei. Bethe and Moliere[10-13] developed a theory describing the angular distribution of charged particles transported through material foils. The angular distribution is the result of many single scatters. This results in an angular distribution that is Gaussian in shape with tails from large angle single and plural scattering. The scattering provides a novel method for obtaining radiographic information with charged particle beams.[14] More recently, scattering information from cosmic ray muons has been shown to be a useful method of radiography for homeland security applications.[15-17]

The dominant part of the multiple scattering polar-angular distribution is Gaussian:

$$\frac{dN}{d\theta} = \frac{1}{2\pi\theta_0^2} e^{-\frac{\theta^2}{2\theta_0^2}},$$

the Fermi approximation, where $\theta$ is the polar angle and $\theta_0$ is the multiple scattering angle, is given approximately by:

$$\theta_0 = \frac{14.1 \text{ MeV}}{p\beta}\sqrt{\frac{l}{X_0}}.$$

The muon momentum and velocity are $p$ and $\beta$, respectively, and $X_0$ is the radiation length for the material. This needs to be convolved with the cosmic ray momentum spectrum in order to describe the angular distribution.

We have approximated the scattering distribution with a model that uses seven momentum groups, $p_i$.

$$\frac{dN}{d\theta} = \sum \frac{A_i}{\sigma_i} e^{-\frac{\theta^2}{2\sigma_i}}$$

$$\sigma_i = \frac{14.1}{p_i}\sqrt{\frac{l}{X_0}}$$

The model has been calibrated to data taken through the three thicknesses of lead described above. We fit the amplitudes, $A_i$, of each energy group, as well as the intrinsic angular resolution and a fixed number of radiation lengths due to the drift tubes and other structural elements of the muon detectors. The model does not account for changes in the shape of the muon spectrum due to stopping. A maximum likelihood fit to the set of lead data is shown in Figure 6. Also shown is the decomposition of one of the data sets into its momentum groups.

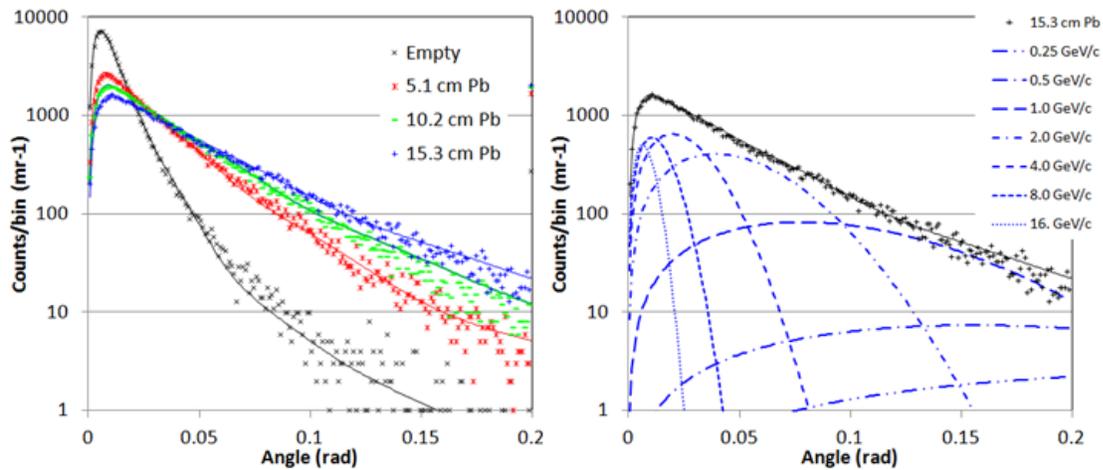

Figure 6) Left) Measured angular distributions for various thickness of lead (points) and the fit (lines) for various thicknesses of lead. Right) The decomposition of the fit into energy groups. Empty shows the angular distribution with no object in the scanner.

The momentum distribution obtained from this fit is compared with previous measurements of the momentum distribution at sea level and a parameterization that has been extrapolated to the altitude at Los Alamos, New Mexico (2231 m) in Figure 7. The agreement is remarkably good above 1 GeV. At lower energies the spectrum is sensitive to stopping and slowing down in the targets and to the electron component of the flux. These effects are not accounted for in this simple model.

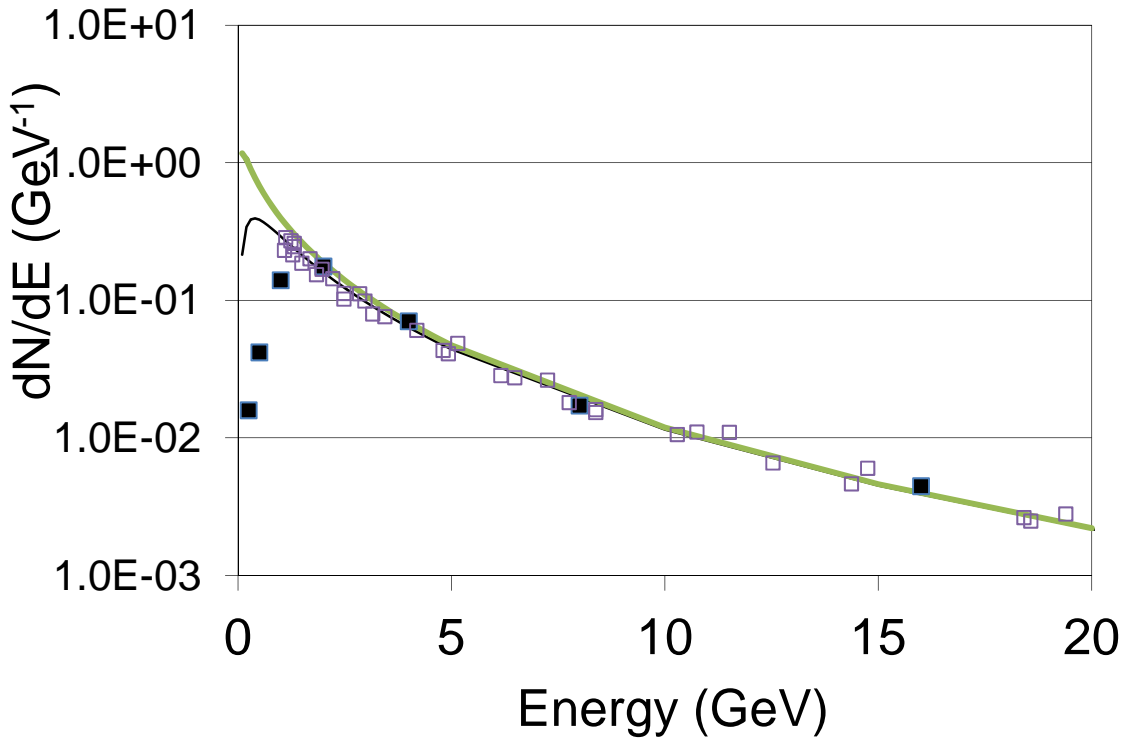

Figure 7) Fitted spectrum (solid symbols) compared to model and previous data (open symbols). Also shown is the extrapolation of the sea level spectrum (black) to the altitude of Los Alamos (green curve).

With the amplitudes fixed by the global fit, described above, a maximum likelihood fit of the angular distribution for each voxel, where $l/X_0$ was the only parameter that was varied, was used to obtain a radiation length image of each of the data sets used above. A composite of the resulting images is shown in Figure 8. The fitted value for the thickness of the test objects in radiation lengths is plotted vs. actual radiation lengths in Figure 5.

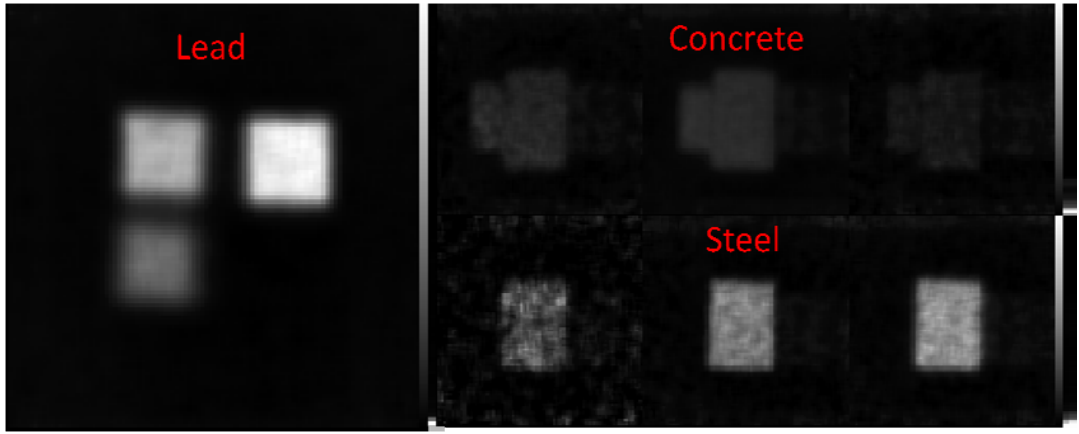

**Figure 8) Radiation length images of the test objects described above.**

The same data have been used to obtain the thickness in both radiation lengths and in attenuation lengths. The known thickness and measured thickness agree to ~10-20%. The experimentally measured radiation lengths and attenuation lengths are plotted vs. each other in Figure 9. The fact that the different materials lie on different curves demonstrates material identification.

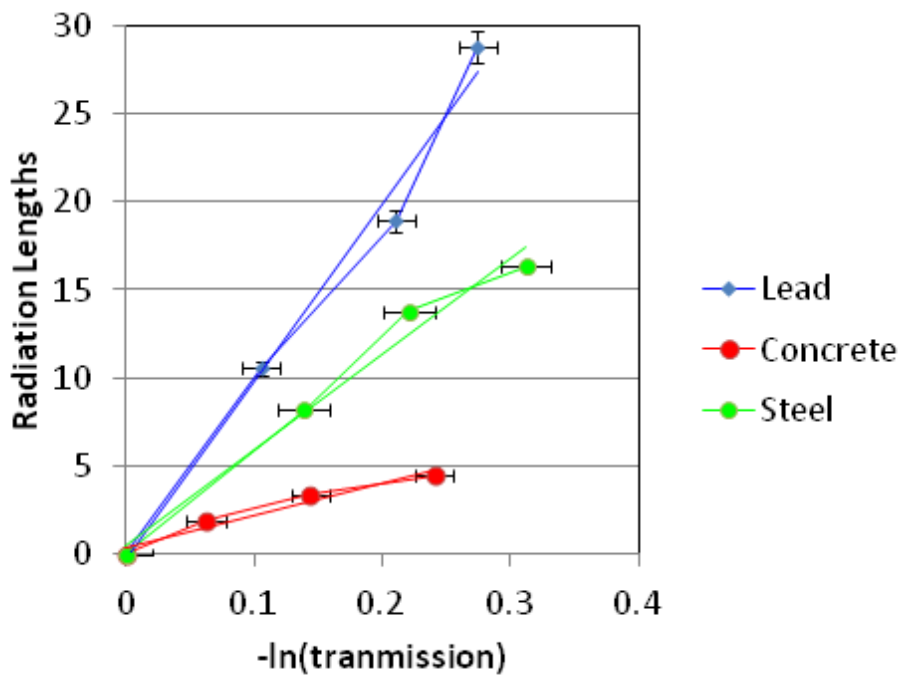

**Figure 9) Radiation lengths vs. attenuation lengths for the different test objects. The different materials lie on lines with different slopes, therefore demonstrating material identification.**

# Conclusions

We have used experimental data collected with the Mini Muon Tracker to validate simple models of cosmic ray muon radiography. The negative of the natural logarithm of transmission is shown to be proportional to material thickness, $t$, for a given material and approximately proportional to $\frac{dE}{dx}t$ across a range of materials. Similarly, we have shown that $l/X_0$, the thickness of the object in radiation length units, can be measured by fitting the polar angular distribution with a set of Gaussian distributions whose amplitudes are fixed at values fitted to calibration data. The momentum distribution fitted to the cosmic ray data is show to agree remarkably well above 1 GeV with previously measured data taken with a magnetic spectrometer. The combination of our measurements provides estimates of both the areal density and the material.

# Acknowledgements

This work was supported in part by the Unites States Department of Energy, LDRD program at LANL and SDRD program at NSTec.

# References

1  E. P. George, Commonwealth Engineer, 455 (1955).
2  L. W. Alvarez, J. A. Anderson, F. Elbedwe*, J. Burkhard, A. Fakhry, A. Girgis, A. Goneid, F. Hassan, D. Iverson, G. Lynch, Z. Miligy, A. H. Moussa, Mohammed, and L. Yazolino, Science **167,** 832 (1970).
3  H. Tanaka, K. Nagamine, N. Kawamura, S. N. Nakamura, K. Ishida, and K. Shimomura, Nuclear Instruments & Methods in Physics Research, Section A (Accelerators, Spectrometers, Detectors and Associated Equipment) **507,** 657 (2003).
4  K. Nagamine, M. Iwasaki, K. Shimomura, and K. Ishida, Nuclear Instruments and Methods in Physics Research Section A: Accelerators, Spectrometers, Detectors and Associated Equipment **356,** 585 (1995).
5  H. Tanaka, K. Nagamine, N. Kawamura, S. N. Nakamura, K. Ishida, and K. Shimomura, Hyperfine Interactions **138,** 521 (2001).
6  K. N. Borozdin, G. E. Hogan, C. Morris, W. C. Priedhorsky, A. Saunders, L. J. Schultz, and M. E. Teasdale, Nature **422,** 277 (2003).
7  K. Nagamine, Proceedings of the Japan Academy, Series B **80,** 179 (2004). F. Grieder, *Cosmic Rays at Earth* (Elsevier Science, 2001).
8  C. Morris, J. W. Hopson, and P. Goldstone, Los Alamos Science **30** (2006).H. A. Bethe, Physical Review **89,** 1256 (1953).
9  P. K. F. Grieder, *Cosmic Rays at Earth* (Elsevier Science, 2001).
10 H. A. Bethe, Physical Review **89,** 1256 (1953).
11 G. R. Lynch and O. I. Dahl, Nuclear Instruments & Methods in Physics Research, Section B (Beam Interactions with Materials and Atoms) **B58,** 6 (1991).
12 G. Moliere, Zeitschrift fur Naturforschung Section A-A Journal of Physical Sciences **2,** 133 (1947).
13 G. Moliere, Zeitschrift fur Naturforschung Section A-A Journal of Physical Sciences **3,** 78 (1948).
14 C. Morris, J. W. Hopson, and P. Goldstone, Los Alamos Science **30** (2006).
15 C. L. Morris, C. C. Alexander, J. D. Bacon, K. N. Borozdin, D. J. Clark, R. Chartrand, C. J. Espinoza, A. M. Fraser, M. C. Galassi, J. A. Green, J. S. Gonzales, J. J. Gomez, N. W. Hengartner, G. E. Hogan,

A. V. Klimenko, M. F. Makela, P. McGaughey, J. J. Medina, F. E. Pazuchanics, W. C. Priedhorsky, J. C. Ramsey, A. Saunders, R. C. Schirato, L. J. Schultz, M. J. Sossong, and G. S. Blanpied, Science and Global Security **16,** 37 (2008).

[16] W. C. Priedhorsky, K. N. Borozdin, G. E. Hogan, C. Morris, A. Saunders, L. J. Schultz, and M. E. Teasdale, Review of Scientific Instruments **74,** 4294 (2003).

[17] L. J. Schultz, G. S. Blanpied, K. N. Borozdin, A. M. Fraser, N. W. Hengartner, A. V. Klimenko, C. L. Morris, C. Oram, and M. J. Sossong, IEEE Transactions on Image Processing **16,** 1985 (2007).